\documentclass[11pt,a4paper,english]{article}

\usepackage{graphicx}
\usepackage{amssymb,latexsym}
\usepackage{amsmath}
\usepackage{epsf}
\usepackage{pdfsync}
\usepackage{epsfig}
\usepackage[parfill]{parskip}
\usepackage[matrix,arrow,color]{xy}
\usepackage[usenames]{color}
\usepackage{hyperref}

\input{xy}
\xyoption{all}

\input{epsf}

\makeatletter

\catcode`@=11
\renewcommand{\section}
{\@startsection{section}{1}{0pt}{\medskipamount}{\medskipamount}{\large\bf}}
\makeatletter\renewcommand{\subsection}
{\@startsection{subsection}{2}{\z@}{-3.25ex plus -1ex minus -.2ex}
{1.5ex plus .2ex}{\it }}

\numberwithin{equation}{section}
\catcode`@=12

\newcommand{\ba}{\begin{eqnarray*}}
\newcommand{\ea}{\end{eqnarray*}}
\newcommand{\ban}{\begin{eqnarray}}
\newcommand{\ean}{\end{eqnarray}}

\newcommand{\IZ}{\mathbb{Z}}
\newcommand{\IC}{\mathbb{C}}
\newcommand{\IP}{\mathbb{P}}

\newcommand{\IR}{\mathbb{R}}

\newcommand{\frM}{\frak{M}}

\newcommand{\cW}{{\cal W}}
\newcommand{\cN}{{\cal N}}
\newcommand{\cM}{{\cal M}}

\newcommand{\cO}{{\cal O}}

\newcommand{\cZ}{{\cal Z}}

\newcommand{\cF}{{\cal F}}

\newcommand{\cG}{{\cal G}}

\newcommand{\mbf}[1]{{\boldsymbol {#1} }}

\def\e{{\,\rm e}\,}

\def\ii{{\,{\rm i}\,}}
\def\dd{{\rm d}}

\def\beq{\begin{equation}}
\def\bee{\begin{equation}}
\def\eeq{\end{equation}}
\def\bea{\begin{eqnarray}}
\def\eea{\end{eqnarray}}
\def\bd{\begin{displaymath}}
\def\ed{\end{displaymath}}

\newcommand{\Cint}{\int\kern-10.5pt-\kern7pt}
\newcommand{\Pint}{{\int\!\!\!\!\!\!-}\,}

\newcommand{\be}{\begin{equation}}
\newcommand{\ee}{\end{equation}}

\newcommand\fverbit{\egroup\item[\fbox{\unhbox\pippobox}]}
\newbox\pippobox

{

\def\be{\begin{equation}}
\def\ee{\end{equation}}
\def\bea{\begin{eqnarray}}
\def\eea{\end{eqnarray}}

\begin{document}

\begin{titlepage}
\setcounter{page}{0}
\begin{flushright}
EMPG--13--08
\end{flushright}

\vskip 1.8cm

\begin{center}

{\Large\bf Refined Chern-Simons theory \\[2mm] and $\mbf{(q,t)}$-deformed
  Yang-Mills theory: \\[3mm] Semi-classical expansion and planar limit}

\vspace{15mm}

{\large\bf Zolt\'an K\"ok\'enyesi$^{(a),(b)}$, Annamaria Sinkovics$^{(b)}$ and
 Richard~J.~Szabo$^{(c)}$}
\\[6mm]
\noindent{\em $^{(a)}$ Institute of Physics\\
Budapest University of Technology and Economics\\ H-1111 Budapest, Budafoki u. 8, Hungary} \\ Email: {\tt kokenyesiz@gmail.com}\\[3mm]
\noindent{\em $^{(b)}$ Institute of Theoretical Physics\\ MTA-ELTE
  Theoretical Research Group \\ E\"otv\"os Lor\'and University \\ 1117
  Budapest, P\'azm\'any s. 1/A, Hungary} \\ Email: \ {\tt
  sinkovics@general.elte.hu}\\[3mm]
\noindent{\em $^{(c)}$ Department of Mathematics\\ Heriot-Watt
  University\\
Colin Maclaurin Building, Riccarton, Edinburgh EH14 4AS, UK\\ 
Maxwell Institute for Mathematical Sciences, Edinburgh, UK\\
The Tait Institute, Edinburgh, UK}\\
Email: \ {\tt
  R.J.Szabo@hw.ac.uk}

\vspace{15mm}

\begin{abstract}
\noindent
We study the relationship between refined Chern-Simons theory on lens
spaces $S^3/\IZ_p$ and $(q,t)$-deformed Yang-Mills theory on the sphere $S^2$.
We derive the instanton partition function of $(q,t)$-deformed $U(N)$ Yang-Mills theory and describe it explicitly as an
analytical continuation of the semi-classical expansion of refined
Chern-Simons theory. The derivations are based on a generalization of
the Weyl character formula to Macdonald polynomials. The expansion is
used to formulate $q$-generalizations of
$\beta$-deformed matrix models for refined Chern-Simons theory, as
well as conjectural formulas for the $\chi_y$-genus of the moduli
space of $U(N)$ instantons on the surface $\cO(-p)\to \IP^1$ for all
$p\geq1$ which enumerate black hole microstates in refined topological
string theory. We study the large $N$ phase structures of the refined
gauge theories, and match them with refined topological string theory on the resolved conifold.
\end{abstract}

\end{center}


\end{titlepage}



\tableofcontents

\bigskip

\section{Introduction}

Refined Chern-Simons theory has been of recent interest because of the
rich structure of the new knot and three-manifold invariants that it
computes, and also because of
its connection to refined topological string theory and the refined
topological vertex~\cite{Aganagic2011,Iqbal:2011kq,Aganagic:2012au,Aganagic:2012hs,Iqbal:2012mt,Nakajima:2012gx,Gorsky:2013jna}. In this paper we
explore its relation to refined topological string theory on the local
Calabi-Yau threefold
\beq
X=\cO(-p)\oplus \cO(p-2) \ \longrightarrow \ \IP^1 \ .
\label{CYfib}\eeq
The usual topological string theory on this
fibration reduces to a $q$-deformation of Yang-Mills
theory on the sphere $S^2$ \cite{Vafa:2004qa,Aganagic:2004js}. The refined
version of this correspondence was studied recently in
\cite{Aganagic2012}, where it was shown that due to the refinement the
reduction leads to a
two-parameter $(q,t)$-deformation of $U(N)$
Yang-Mills theory on $S^2$.

The focus of the present work is on this  $(q,t)$-deformed
two-dimensional gauge theory. Starting with the refined version of the
heat kernel expansion for the partition function of the theory, we
show that it can be transformed in various insightful ways. From a
three-dimensional perspective, we demonstrate explicitly that the
$(q,t)$-deformed gauge theory defines an analytic continuation of
refined Chern-Simons theory on the lens space $L(p,1)= S^3/\IZ_p$.
Via Poisson resummation we obtain the instanton expansion of the
two-dimensional gauge theory, and also derive new $\beta$-deformed matrix
model representations of the partition function.

The corresponding refined topological string partition function can also be studied by counting 
BPS bound states of D4-D2-D0 branes in $X$ with angular momentum and R-charge, which is computed by the
Hirzebruch $\chi_y$-genus of the moduli space of $U(N)$ instantons in
a topologically
twisted ${\cal N} =4$ gauge theory on the worldvolume $D= \cO(-p)\to
\IP^1$ of the D4-branes. Using our expansions of the two-dimensional
gauge theory partition functions, we present new conjectural formulas
for the contributions of curve classes to the $\chi_y$-genus for arbitrary $p$. 

We also describe the planar large $N$ limit and the phase structure of
the $(q,t)$-deformed two-dimensional Yang-Mills theory, both directly
in terms of the heat kernel expansion
and also from the point of view of the instanton expansion. We find an
analogous phase behaviour to the unrefined $q$-deformed
case~\cite{Arsiwalla:2005jb,Jafferis:2005jd,Caporaso2005}: The theory
exhibits the usual phase transition which, as a function of the area
$A = \tau_2 \, p$ where $\tau_2$ is identified with the 't Hooft
coupling, occurs only for $p > 2$. We identify the critical curve and
compute the free energy explicitly in the small area phase. Using this
result we discuss the connection between the refined gauge theory and the emergent conifold geometry in
refined topological string theory.

It would be interesting to find
other large $N$ limits, for instance involving a limit where one of
the equivariant rotations vanishes as was studied in~\cite{Aganagic:2011mi} in the context
of refined topological string theory, but we have not found any other
planar limit which leads to sensible results. To explore truly
non-trivial effects of the $\beta$-deformation, one should proceed to
construct the $\frac1N$-expansion of the theory based on a suitable
double scaling limit; this has been studied in~\cite{Brini:2010fc} for
the $\beta$-deformation of the Chern-Simons matrix model, and it would
be interesting to extend their results to include the full quantum
group $\beta$-deformation appropriate to the refined Chern-Simons
matrix model.
It would also be
interesting to understand the implications of this phase structure to the
duality between $(q,t)$-deformed two-dimensional Yang-Mills theory and
four-dimensional $\cN = 2$ gauge theories with two superconformal fugacities~\cite{Gadde:2011uv,Tachikawa:2012wi}.

The structure of the paper is as follows. In \S\ref{genasp} we
discuss the formalism of the $(q,t)$-deformation of two-dimensional Yang-Mills theory, and the semiclassical expansion of its partition function.
In \S\ref{RefCS} we describe the connection to refined
Chern-Simons theory on the Lens space $S^3/\IZ_p$, and its  formulation in terms of $\beta$-deformed matrix models.
In \S\ref{chiy} we discuss the relationship between the
two-dimensional gauge theory and the refined black hole partition
function which enumerates spinning D4-D2-D0 brane bound states, and
present the conjectural formulas for the $\chi_y$-genus. In
\S\ref{Planar} we study the planar limit and phase structure of the
$(q,t)$-deformed two-dimensional Yang-Mills theory, while
\S\ref{geometry} contains final remarks concerning the connection to
the emergent geometry in the large $N$ limit. Appendix~A summarises
the main features of the generalised Weyl denominator formula that is
used in the main text to derive the pertinent semi-classical expansions. 

\section{$(q,t)$-deformed Yang-Mills theory on $S^2$}  \label{genasp}

\subsection{General aspects}

The partition function for the $(q,t)$-deformation of $U(N)$
Yang-Mills theory on the sphere $S^2$ can be written as a
generalization of the Migdal heat kernel expansion given by~\cite{Aganagic2012}
\bea
Z(q,t,Q;p) = \sum_R\, \frac{\dim_{q,t}(R)^2}{g_R}\ q^{\frac p2\,
  (R,R)}\, t^{p\,(\rho,R)}\, Q^{|R|} \ ,
\label{ZqtQpR}\eea
where the sum runs over all irreducible unitary representations of
the $U(N)$ gauge group which are parametrized by partitions
$R=(R_1,\dots,R_N)$, with $R_1\geq R_2\geq\cdots \geq R_N\geq 0$, such
that $R_i$ is the length of the $i$-th row of the corresponding Young
diagram.
Here the deformation parameters are
\bea
q=\e^{-\epsilon_1} \qquad \mbox{and} \qquad t=\e^{-\epsilon_2} \ , 
\eea
where $(\epsilon_1,\epsilon_2)$ may be regarded
as parameterizing either the
$\Omega$-background in the corresponding five-dimensional gauge
theory, the left/right angular momentum of BPS states of spinning
M2-branes in M-theory, or the strength of the non-selfdual graviphoton
background in topological string theory. For simplicity of presentation, below we shall write some formulas for the
case when the refinement parameter
\bea
\beta=\frac{\epsilon_2}{\epsilon_1}
\eea
is a positive integer, and then analytically continue final results to
arbitrary $\beta\in\IR$.
The parameter $Q=\e^{-\ii\theta}$ is the contribution from
the two-dimensional theta-angle. 
For a pair of weights
$\lambda=(\lambda_1,\dots,\lambda_N)$ and
$\lambda'=(\lambda_1',\dots,\lambda_N')$, we define
\bea
(\lambda,\lambda'\,)=\sum_{i=1}^N\, \lambda_i\, \lambda_i' \ ,
\eea
and
\beq
\rho_i=\mbox{$\frac12$} \, (N+1-2i)
\eeq
for $i=1,\dots,N$ are the components of the Weyl vector $\rho$ for
$U(N)$. We shall often assume that the rank $N$ is
odd, so that $\rho\in\IZ^N$; this restriction is not necessary but it
will simplify some of our analysis below. The quantity
\bea
|R|=\sum_{i=1}^N\, R_i
\eea
is the total number of boxes in the Young diagram associated to the
representation $R$. 
The $(q,t)$-deformed dimension of the representation $R$ is
\bea
\dim_{q,t}(R)=\prod_{m=0}^{\beta-1} \ \prod_{1\leq i<j\leq N}\,
\frac{\big[R_i-R_j+\beta\,(j-i)+m\big]_q}{\big[\beta\, (j-i)+m\big]_q}
\ ,
\label{qtdim}\eea
where
\bea
[x]_q=\frac{q^{x/2}-q^{-x/2}}{q-q^{-1}}
\label{qnumber}\eea
for $x\in\IR$ is a $q$-number. The Macdonald inner product normalization is given by
\bea
g_R=\prod_{m=0}^{\beta-1} \ \prod_{1\leq i<j\leq N}\,
\frac{\big[R_i-R_j+\beta\,(j-i)+m\big]_q}{\big[R_i-R_j+ \beta\, (j-i)-
  m\big]_q} \ .
\eea
Our normalization differs from that
of~\cite{Aganagic2012}, wherein the refined quantum dimensions are
multiplied by the factor
\bea
S_{00} = \prod_{m=0}^{\beta-1}\ \prod_{i=1}^{N-1}\, \big(q^{-m/2}\,
t^{-i/2}-q^{m/2}\, t^{i/2}\big)^{N-i} \ .
\label{S00}\eea
This normalization factor will become important later on in our
comparisons with refined Chern-Simons theory and its connection to
refined topological string theory.
We have also taken a different normalization for the $q$-number
(\ref{qnumber}) such that $[x]_q=x+O(\log q)$ in the limit $q\to1^-$;
the present normalization is more useful for considering various limits below.

For our computations below we will require an explicit expression for the
partition function (\ref{ZqtQpR}) in terms of highest weight variables. For this, we
define shifted weights $n_i$ by
\bea
n_i=R_i+\beta\, \rho_i
\label{nRbetarho}\eea
for $i=1,\dots,N$. The range of these integers is
$+\infty>n_1>n_2>\cdots >n_N>-\infty$. We can use the Weyl reflection
symmetry of the summand of the
partition function (\ref{ZqtQpR}) to remove the restriction to the fundamental chamber of the summation over
$n=(n_1,\dots,n_N)$, and assume $n_i\neq n_j$ for all
$i\neq j$. Then we can extend the summation range over all of
$n\in\IZ^N$ using the fact that $[n_i-n_j]_q=0$ whenever $n_i=n_j$. Up
to overall normalization, the partition function (\ref{ZqtQpR}) can thus be written
as a quantum $\beta$-deformation of the discrete Gaussian matrix model given by
\bea
Z(q,t,Q;p)= \sum_{n\in\IZ^N}\,
\Delta_{q,t}(\epsilon_1\, n)\,\Delta_{q,t}(-\epsilon_1\, n) \, \e^{-\frac{p\,\epsilon_1}2\,
  (n,n)-\ii\theta\, |n| } \ ,
\label{ZqtQpn}\eea
where the Macdonald measure is given by
\bea
\Delta_{q,t}(x)=\prod_{m=0}^{\beta-1} \ \prod_{1\leq i<j \leq N}\,
\big(q^{-m/2}\, \e^{(x_j-x_i)/2}-q^{m/2}\, \e^{(x_i-x_j)/2} \big)
\label{Macmeasure}\eea
for $x=(x_1,\dots,x_N)\in\IC^N$. In the unrefined limit $\beta=1$, the
measure (\ref{Macmeasure}) reduces to the usual Weyl determinant
\beq
\Delta(x)=\Delta_{q,q}(x) = \prod_{1\leq i<j\leq N}\, 2\sinh\Big(\,
\frac{x_i-x_j}2\, \Big)
\eeq
which arises as a $q$-deformation of the Vandermonde determinant.

An interesting limit of this model is the one in which we send
$\epsilon_1,\epsilon_2 \to0$ and $p\to\infty$ with the parameters
$\beta$ and
\bea
a= \epsilon_1\, p
\label{aepsilon1p}\eea
held fixed. In that case, all $q$-numbers reduce smoothly to ordinary numbers
and the $(q,t)$-deformed Yang-Mills theory (\ref{ZqtQpR}) reduces to a
$\beta$-deformation of ordinary Yang-Mills theory given by
\bea
\cZ_N^{{\rm YM},\beta}(a,\theta) = \sum_R\ \prod_{m=0}^{\beta-1}\
\prod_{1\leq i<j\leq N}\, \Big(\big(R_i-R_j+\beta\,
(j-i)\big)^2-m^2\Big) \ \e^{-\frac{a}2\,(R,R+2\, \beta\, \rho)}\,
\e^{-\ii\theta\, |R|} \ ,
\label{ZNYMbetaR}\eea
which for $\beta=1$ coincides with ordinary (undeformed, unrefined)
$U(N)$ Yang-Mills theory on the sphere $S^2$. As we discuss in \S\ref{RefCS}, from a three-dimensional
perspective the $(q,t)$-deformed
gauge theory defines an analytical continuation of refined
Chern-Simons theory~\cite{Aganagic2011} on the lens space $L(p,1)= S^3/\IZ_p$ to arbitrary values
of the Chern-Simons level $k$; the limit $p\to\infty$ of infinite
degree of the Seifert fibration $S^3/\IZ_p\to S^2$ thereby reduces the
partition function of refined Chern-Simons theory to that of the
$\beta$-deformation (\ref{ZNYMbetaR}) of ordinary Yang-Mills
theory. Alternatively, we can take this limit directly in
(\ref{ZqtQpn}) to obtain a $\beta$-deformation of the discrete
Gaussian matrix model
\bea
\cZ_N^{{\rm YM},\beta}(a,\theta) = \sum_{n\in\IZ^N}\
\prod_{m=0}^{\beta-1}\ \prod_{1\leq i<j\leq N}\, \big((n_i-n_j)^2 - m^2\big) \ \e^{-\frac{a}2\, (n,n) -\ii\theta\,
 |n|} \ .
\label{ZNYMbetan}\eea
From the point of view of topological string theory, the limit
$p\to\infty$ should be understood as a singular limit of the
underlying Calabi-Yau geometry (\ref{CYfib}). In
this setting the partition function (\ref{ZNYMbetan}) defines a discrete version of the $\beta$-deformed Gaussian
matrix ensemble considered in~\cite{DijkgraafVafa2009}; in \S\ref{geometry} we show that the planar limit
coincides with refined topological string theory on the conifold
represented as the $c=1$ string theory at a non-selfdual radius.

\subsection{Semi-classical expansion}

We will now derive the dual description of the refined $q$-deformed gauge
theory in terms of instanton degrees of freedom, which is provided by performing a
modular inversion of the series (\ref{ZqtQpn}). For this, let us
rewrite this series in the form
\bea
Z(q,t,Q;p)= \sum_{n\in\IZ^N}\, \Delta(-\epsilon_1\,n)\,
\widetilde{\Delta}_{q,t}(\epsilon_1\, n)\,\e^{-\frac{p\,\epsilon_1}2\,
  (n,n)-\ii\theta\, |n| } \ ,
\eea
where
\bea
\widetilde{\Delta}_{q,t}(x):=\frac{\Delta_{q,t}(x)\,
  \Delta_{q,t}(-x)}{\Delta(-x)} \ .
\eea
Substituting the (generalised) Weyl denominator
formulas for $\Delta(x)$ and $\widetilde{\Delta}_{q,t}(x)$ from Appendix~A, after some simple manipulations and dropping
of overall normalisations throughout we can recast the partition
function in the form
\bea
Z(q,t,Q;p) &=& \sum_{w\in S_N}\, \varepsilon(w) \ \sum_{n\in\IZ^N}\, \e^{-\frac{p\,\epsilon_1}2\,
  (n,n)-\ii\theta\, |n| } \, \e^{\epsilon_1\, (w(\rho)-\beta\,
  \rho,n)} \nonumber \\ && \times \ \sum_{\mu \in\Lambda_\beta} \ \sum_{w'\in
  S_N} \, \e^{-\epsilon_1\,
  (\mu ,w'{}^{-1}(n))}\ \Pi_\mu \big(\beta\, w\,w'(\rho);q,t\big) \ .
\eea
The Poisson resummation of this series is now accomplished through an
elementary Gaussian integration, and one finds
\bea
Z(q,t,Q;p)= \sum_{m \in\IZ^N}\, \e^{-\frac{2\pi^2}{p\,\epsilon_1}\,
  (m,m)-\frac{2\pi\,\theta}{p\, \epsilon_1}\, |m|} \ \cW_{q,t}(p;m)
\label{ZqtQpm}\eea
where
\bea
\cW_{q,t}(p;m) &=& \sum_{w\in S_N}\, \varepsilon(w)\, \e^{\frac{2\pi\ii}p\,
(m,w(\rho)-\beta\, \rho)} \, \e^{-\frac{\beta\,\epsilon_1}p\,
(w(\rho),\rho)} \label{Wcm} \\ && \times\ \sum_{\mu \in\Lambda_\beta} \ \sum_{w'\in S_N}\,
\e^{-\frac{2\pi\ii}p\, (m,w'(\mu ))}\,
\e^{\frac{\epsilon_1}{2p}\, (\mu ,\mu -2w'{}^{-1}\, (w(\rho)-\beta\,
  \rho))}\ \Pi_\mu \big(\beta\,w\,w'(\rho);q,t\big) \ . \nonumber 
\eea

To understand the meaning of this series, we note that at the classical level the refined
two-dimensional gauge
theory is just ordinary Yang-Mills theory on the sphere $S^2$~\cite{Aganagic2012}. Using a
gauge transformation, we can conjugate Yang-Mills connections of a
$U(N)$ gauge bundle over $S^2$ so that they are valued in the Lie
algebra of the maximal torus $U(1)^N\subset U(N)$; they
correspond to sums of $U(1)$ Dirac monopole connections with
topological charges $m_i\in\IZ$ for $i=1,\dots,N$. The classical
Yang-Mills action with theta-angle evaluated on such a configuration is given by
\bea
S_N^{\rm YM}(a,\theta;m)=\frac{2\pi^2}{a}\, \sum_{i=1}^N\,
\Big(\, m_i^2+\frac{\theta\, m_i}{\pi}\Big) \ ,
\eea
where $a$ is the dimensionless Yang-Mills coupling constant on
$S^2$. With the identification (\ref{aepsilon1p}), we see that the exponential prefactors in the series (\ref{ZqtQpm}) have a
natural interpretation as the
classical contributions $\e^{-S_N^{\rm YM}(a,\theta; m)}$ of instantons to the refined gauge theory path
integral, while the sums (\ref{Wcm}) are the fluctuation
determinants around each instanton. 

Note that the residual Weyl
symmetry $S_N$ of the $U(N)$ gauge group after conjugation to the
maximal torus permutes the different
components of the classical monopole solutions. The classical field theory is invariant under this residual gauge symmetry.
However, the path integral measure defining
the quantum gauge theory differs from that of the unrefined case and
is essentially determined by the Macdonald measure
(\ref{Macmeasure})~\cite{Aganagic2012}; this is reflected in the form
of the quantum fluctuations $\cW_{q,t}(p;m)$ which are not invariant
under all
gauge transformations in the Weyl group $S_N$. Whence the semi-classical expansion of the $\beta$-deformation of $q$-deformed Yang-Mills
theory explicitly breaks a discrete part of the gauge symmetry; this is
due to the way in which the quantum group nature of the gauge symmetry
is manifested in
the refined case which typically requires a notion of ``twisted'' invariance~\cite{Szabo2013}. In the following we will find several interesting
consequences of this symmetry breaking.

\section{Refined Chern-Simons theory on $S^3/\IZ_p$} \label{RefCS}

\subsection{Semi-classical expansion}

We shall now describe the precise sense in which the $(q,t)$-deformed
gauge theory on $S^2$ is an analytic continuation of the refinement of
Chern-Simons theory on the lens space $S^3/\IZ_p$ defined
in~\cite{Aganagic2011,Aganagic2012}.

An expression for the path integral of $U(N)$ refined Chern-Simons
gauge theory on
$S^3/\IZ_p$ at level $k\in\IZ$ is derived
in~\cite{Aganagic2011} using cutting and gluing rules of
three-dimensional topological quantum field theory. The field theory depends
on the parameters
\beq
q= \e^{g_s} \qquad \mbox{and} \qquad t=q^\beta
\eeq
defined in terms of the genus expansion parameter
\bea
g_s:=\frac{2\pi\ii}{k+\beta\,N} \ .
\label{gsepsilon1}\eea
The
partition function is~\cite[\S5]{Aganagic2011}
\beq
\cZ_N^{{\rm CS},\beta}(g_s;p) = \sum_R\, (T_R)^p\  g_R^{-1}\,
(S_{0R})^2 \ ,
\label{ZNCSdef}\eeq
where (up to overall normalization) $S_{0R}=S_{00}\, \dim_{q,t}(R)$
and
\beq
T_R=q^{\frac12\, (R,R)}\ t^{(R,\rho)} \ .
\eeq

The series (\ref{ZNCSdef}) is formally identical to (\ref{ZqtQpR}), except
that now the summation is finite and restricted to the integrable
representations of $U(N)$ at level $k\in\IZ$. 
With the same redefinition of
weight vectors (\ref{nRbetarho}), we can write the sum over integrable
representations as a sum over $n\in\IZ^N_{k+\beta\,N}$. Using the
generalised Weyl denominator formulas from Appendix~A, we can then
write the refined Chern-Simons partition function as a lattice Gauss
sum
\bea
\cZ_N^{{\rm CS},\beta}(g_s;p) &=& \sum_{w\in S_N}\, \varepsilon(w) \ \sum_{n\in\IZ_{k+\beta\,N}^N}\,
\e^{\frac{\pi\ii p}{k+\beta\,N} \,
  (n,n)} \, \e^{\frac{2\pi\ii}{k+\beta\,N}\, (w(\rho)-\beta\,
  \rho,n)} \nonumber \\ && \times \ \sum_{\mu \in\Lambda_\beta} \ \sum_{w'\in
  S_N}\, \e^{\frac{2\pi\ii}{k+\beta\,N} \,
  (\mu ,w'{}^{-1}(n))}\ \Pi_\mu \big(\beta\, w\,w'(\rho);q,t\big) \ .
\eea
We now apply the quadratic reciprocity formula for Gauss sums to
rewrite the sum over $n\in\IZ^N_{k+\beta\,N}$ as a sum over
$r\in\IZ^N_p$, and again dropping irrelevant overall normalization
factors we find
\bea
\cZ_N^{{\rm CS},\beta}(g_s;p) =\sum_{r\in\IZ_p^N}\, \cZ_N^{{\rm
    CS},\beta}(g_s;p;r) :=  \sum_{r\in\IZ_p^N}\,
\e^{-\frac{\pi\ii(k+\beta\,N)}{p}\, (r,r)}\ \cW_{N,k}^\beta(p;r)
\label{ZNCSbetasemiclass}\eea
where
\bea
\cW_{N,k}^\beta(p;r)&=& 
\sum_{w\in S_N}\, \varepsilon(w)\, \e^{-\frac{2\pi\ii}p\,
  (r,w(\rho)-\beta\,\rho)}\, \e^{\frac{2\pi\ii\beta}{p\,(k+\beta\,N)} \, (w(\rho),\rho)}
\label{ZNCSbetafluct} \\
&& \times\ \sum_{\mu \in\Lambda_\beta} \ \sum_{w'\in S_N}\, \e^{\frac{2\pi\ii}p\,
  (r,w'(\mu ))}\, \e^{-\frac{\pi\ii}{p\,(k+\beta\,N)}\,
  (\mu ,\mu -2w'{}^{-1}(w(\rho)-\beta\,\rho))}  \ \Pi_\mu \big(\beta\,
w\,w'(\rho);q,t\big) \ . \nonumber
\eea

To understand the meaning of the sum (\ref{ZNCSbetasemiclass}), we
note that at the classical level the refined gauge theory is
identical to ordinary Chern-Simons theory on $S^3/\IZ_p$~\cite{Aganagic2011}. The
critical points of the Chern-Simons action functional are flat connections.
Gauge equivalence classes of flat $U(N)$ connections
on the lens space $S^3/\IZ_p$ 
are in one-to-one correspondence with isomorphism classes of
$N$-dimensional unitary representations of the fundamental group
$\pi_1(S^3/\IZ_p)=\IZ_p$. Using a gauge transformation, any such
representation can be taken to have image in the maximal torus
$U(1)^N\subset U(N)$. Thus the exponential prefactor
in (\ref{ZNCSbetasemiclass}) is easily identified as $\e^{-S_{N,k}^{{\rm CS},\beta}(p;r)}$, where
\bea
S_{N,k}^{{\rm CS},\beta}(p;r) = \frac{\pi\ii (k+\beta\,N)}p\, \sum_{i=1}^N\, r_i^2
\label{SNCSr}\eea
is the value of the classical Chern-Simons action on the lens space $S^3/\IZ_p$
at the flat connection parameterized by the torsion vector
$r=(r_1,\dots,r_N)\in\IZ_p^N$~\cite{Griguolo2006}. The expansion
(\ref{ZNCSbetasemiclass}) is then evidently the
semi-classical expansion of the refined Chern-Simons gauge theory,
with the sums (\ref{ZNCSbetafluct}) representing the one-loop quantum fluctuation determinants about
the classical solutions. In the unrefined limit $\beta=1$, the second line of
(\ref{ZNCSbetafluct}) is equal to one (see Appendix~A) and the
expression (\ref{ZNCSbetasemiclass}) coincides with the semi-classical
formula for
the ordinary $U(N)$ Chern-Simons gauge theory partition function on
$S^3/\IZ_p$~\cite{Griguolo2006}.

By gauge
invariance, the classical
Chern-Simons action is invariant under the residual gauge symmetry
generated by the action of the Weyl group of $U(N)$ which permutes the
different components.
On the other hand, as before the $\beta$-deformation breaks
this gauge symmetry. Hence the refinement of Chern-Simons theory is
generically sensitive to
gauge equivalent flat connections which are related by a discrete
gauge transformation in the subgroup $S_N$. 

At the classical level, the equivalence between Chern-Simons theory on
$S^3/\IZ_p$ in the background of flat connections parametrized by
$\IZ_p^N$ and Yang-Mills theory on $S^2$ in the background of
two-dimensional instantons parametrized by $\IZ^N$ is well-known~\cite{Caporaso2005,Griguolo2006}: Every flat connection on $S^3/\IZ_p$ is
the pullback by the bundle projection of the
Seifert fibration $S^3/\IZ_p\to S^2$ of a configuration of Dirac monopoles on the sphere
$S^2$ with magnetic charges $m_i$, and the holonomy of this abelian gauge
connection depends only on the values of the monopole numbers $m_i$
modulo~$p$. To state the equivalence at the quantum level, we note
that the fluctuation factors (\ref{Wcm}) are quasi-periodic in $m$ with period
$p$ in the sense that
\beq
\cW_{q,t}(p;m+p\, n) = \e^{-2\pi\ii(\beta-1)\, (n,\rho)} \ \cW_{q,t}(p;m)
\label{quasiperiodic}\eeq
for all $n\in\IZ^N$. It is thus natural to factorise the fluctuations by decomposing the
sum over multi-monopole charges $m\in\IZ^N$ as $m_i=p\, n_i+r_i$, where $n_i\in\IZ$ and
$r_i\in\IZ_p$ for $i=1,\dots,N$. Then the instanton expansion (\ref{ZqtQpm}) becomes
\bea\label{ZqtQptorsion}
Z(q,t,Q;p) &=& \sum_{r\in\IZ_p^N} \, \cZ_N^{{\rm
    CS},\beta}(-\epsilon_1;-p;r) \\ && \times
\ \sum_{n\in\IZ^N} \, \e^{-\frac{2\pi^2\,
  p}{\epsilon_1}\, (n,n)-\frac{2\pi\,\theta}{p\,\epsilon_1}\,
|r+p\,n|-\frac{4\pi^2}{\epsilon_1}\, (r,n)}\, \e^{-2\pi\ii(\beta-1)\,
(n,\rho)} \ . \nonumber
\eea
The meaning of the additional terms from the sum over $n\in\IZ^N$ in (\ref{ZqtQptorsion}) will be elucidated in~\S\ref{chiy}.

\subsection{Matrix models}

In \S\ref{Planar} we will treat the planar limit of the
instanton partition function; for this, it is more convenient to have
available a matrix integral representation of the refined Chern-Simons
fluctuation terms (\ref{ZNCSbetafluct}). Such a representation can be
achieved by performing the Poisson resummation of the original
series (\ref{ZqtQpn}) in a different way to write the modular
inversion as
\bea
Z(q,t,Q;p)=\sum_{m\in\IZ^N}\, \int_{\IR^N}\ \prod_{i=1}^N\, \dd x_i \
\e^{2\pi\ii m_i\, x_i-\ii\theta\, x_i}\ \cF_q(p;x)\, \cF_{q,t}(p;x) \ ,
\label{ZqtQpmodular}\eea
where
\bea
\cF_{q,t}(p;x):= \widetilde{\Delta}_{q,t}(\epsilon_1\,x)\, \e^{-\frac{p\,\epsilon_1}4\,
  |x|^2}
\eea
and $\cF_q(p;x):=\cF_{q,q}(p;-x)$.
By using the ordinary Weyl denominator formula one
shows~\cite{Arsiwalla:2005jb,Caporaso2005} that the Fourier transform
of the function $\cF_q(p;x)$ is given by
\bea
\widehat{\cF}_q(p;m):= \int_{\IR^N}\ \prod_{i=1}^N\, \dd x_i \ \e^{2\pi\ii
  m_i\, x_i} \ \cF_q(p;x) = \Delta\big(-\mbox{$\frac{4\pi\ii}p$}\,
m\big)\, \e^{-\frac{4\pi^2}{p\, \epsilon_1}\, |m|^2} \ .
\eea
On the other hand, using the generalised Weyl denominator formula from
Appendix~A we compute the Fourier transformation
\bea
\widehat{\cF}_{q,t}(p;m)&=& \e^{-\frac{4\pi^2}{p\, \epsilon_1}\, |m|^2}\
\sum_{w\in S_N}\, \varepsilon(w)\, \e^{\frac{4\pi\ii
    \beta}p\, (w(\rho),m)} \nonumber \\ && \times \ \sum_{\mu
  \in\Lambda_\beta}\, \e^{\frac{4\pi\ii}p\, (\mu ,m)}\, \e^{\frac{\epsilon_1}p\, (\mu ,\mu +2\beta\, w(\rho))} \ \Pi_\mu \big(\beta\,
w(\rho);q,t\big) \ .
\eea
Using the standard convolution formula to evaluate the product Fourier
transformation in (\ref{ZqtQpmodular}), we thus find that the
fluctuation terms in (\ref{ZqtQpm}) can be alternatively represented
in the form of a matrix integral
\bea
\cW_{q,t}(p;m) = \int_{\IR^N} \ \prod_{i=1}^N\, \dd
  u_i \ \e^{-\frac{2\pi^2}{p\, \epsilon_1}\, u_i^2} \
  \Delta\big(\mbox{$\frac{2\pi\ii}p$}\, (
u-m) \big)\,\widehat{\Delta}_{q,t,p}\big(\mbox{$\frac{2\pi\ii}p$}\,
(u+m)\big)
\label{cWm}\eea
where
\bea
\widehat{\Delta}_{q,t,p}(x):=\sum_{w\in S_N}\, \varepsilon(w)\,
\e^{\beta\,(w(\rho),x)} \ \sum_{\mu \in\Lambda_\beta}\, \e^{(\mu ,x)}\,
\e^{\frac{\epsilon_1}p\, (\mu ,\mu +2\beta\, w(\rho))} \ \Pi_\mu \big(\beta\,
w(\rho);q,t\big) \ .
\label{hatDeltaqt}\eea
Note that in the unrefined limit $\beta=1$, only the $\mu =0$ term
contributes in the second sum of (\ref{hatDeltaqt}) with
$\Pi_0(\lambda;q,t)=1$ (see Appendix~A), so that
$\widehat{\Delta}_{q,q,p}(x)=\Delta(x)$ and (\ref{cWm}) coincides with
the standard fluctuation integral of the unrefined $q$-deformed gauge
theory on $S^2$~\cite{Arsiwalla:2005jb,Caporaso2005}. With the
rescalings $u_i\to \ii u_i/2\pi$ together with the analytic
continuation $g_s=-\epsilon_1$ of
the genus expansion parameter (\ref{gsepsilon1}), we identify
(\ref{cWm}) as a $\beta$-deformed matrix model for refined
Chern-Simons gauge theory on the lens space $S^3/\IZ_p$. For $p=1$ and
$\beta\in\IZ_{>0}$ a
similar matrix integral is obtained in~\cite{Aganagic2011}; the
equivalence between the discrete matrix model (\ref{ZqtQpn}) and the
$\beta$-deformed Stieltjes-Wigert matrix models for refined
Chern-Simons theory is proven in~\cite{Szabo2013}. 

From (\ref{cWm}) we can equivalently cast the refined Chern-Simons partition
function in the form of a $\beta$-deformed unitary matrix model. For this, we rescale $u\to p\,
u$ and use quasi-periodicity (\ref{quasiperiodic}) of the measure
factor $\Delta\big(\mbox{$\frac{2\pi\ii}p$}\, (p\,u-m)\big)\,\widehat{\Delta}_{q,t,p}\big(\mbox{$\frac{2\pi\ii}p$}\,
(p\, u+m)\big)$ under
integer translations of the integration variables $u\in\IR^N$. Hence
we can truncate the integration domain to $u\in [0,1)^N$ by summing over
all integer shifts of $u_i$, which we do for each $i=1,\dots,N$ by using the modular
transformation
\bea
\sum_{n_i\in\IZ}\, \e^{-\frac{2\pi^2\, p}{\epsilon_1}\, (u_i-n_i)^2
  -2\pi\ii(\beta-1)\, n_i\, \rho_i} &=&
\sqrt{\frac{\epsilon_1}{2\pi\,p}} \ \e^{\pi\ii(\beta-1)\, \rho_i\,
  u_i+\frac{\epsilon_1}{2p}\, (\beta-1)^2\, \rho_i^2}\nonumber \\ &&
\times \ 
\vartheta_3\big(\mbox{$\frac{\ii\epsilon_1}{2\pi\, p}$} \,,\, 2\pi\, 
u_i-\mbox{$\frac{\ii\epsilon_1}p$}\, (\beta-1)\, \rho_i \big)
\eea
of the Jacobi-Erderlyi elliptic function
\beq
\vartheta_3(\tau,z):= \sum_{n\in\IZ}\, \e^{\pi\ii n^2\,\tau + \ii
  n\,z} \ .
\label{Jacobitheta}\eeq
Dropping irrelevant overall constants and rescaling
$u_i=\frac{\phi_i}{2\pi}$ with $\phi_i\in[0,2\pi)$ for $i=1,\dots,N$,
we can rewrite the fluctuation factors (\ref{cWm}) in the form of
compact angular integrations
\bea
\cW_{q,t}(p;m) &=& \int_{[0,2\pi)^N}\ \prod_{i=1}^N\,
\frac{\dd\phi_i}{2\pi} \ \e^{\frac\ii2\, (\beta-1)\, \rho_i\,
  \phi_i}\
\vartheta_3\big(\mbox{$\frac{\ii\epsilon_1}{2\pi\, p}$} \,,\,
\phi_i-\mbox{$\frac{\ii\epsilon_1}p$}\, (\beta-1)\, \rho_i\big)
\nonumber \\ && \qquad \qquad \qquad \qquad \times \
\Delta\big(\ii\phi - \mbox{$\frac{2\pi\ii m}p$}
\big)\,\widehat{\Delta}_{q,t,p}\big(\ii\phi+ \mbox{$\frac{2\pi \ii m}p$}\big) \ .
\label{cWmunitary}\eea
For $p=1$ and $\beta\in\IZ_{>0}$, a similar unitary matrix model is given
in~\cite[Appendix~B]{Aganagic2011}. 

\section{Refined black hole entropy and the $\chi_y$-genus} \label{chiy}

Let us now discuss the precise relationship between the
$(q,t)$-deformed Yang-Mills theory on $S^2$ and refined black hole
partition functions in four dimensions. According
to~\cite{Aganagic2012}, refined black hole degeneracies are obtained
by computing
a protected spin character in four dimensions which enumerates spinning bound states of
D4-D2-D0 brane systems with $N$ D4-branes wrapped on the divisor $D= \cO(-p)\to
\IP^1$ inside the ambient Calabi-Yau threefold (\ref{CYfib}), and D2-branes wrapping the base $\IP^1$. The BPS
degeneracies in this case are computed by the 
$\chi_y$-genus of the moduli space of $U(N)$ instantons in a
topologically twisted $\cN=4$ gauge theory on the D4-brane
worldvolume $D$. Then the black hole partition function is given by
\bea
\cZ_N^{{\rm BH}}(\phi_0,\phi_2,y;p) = \sum_{n,c\in\IZ}\,
\e^{-\phi_0\, n-\phi_2\, c}\ \chi_y\big(\frM_{n,c}(\cO(-p)\big) \ ,
\eea
where
\bea
\chi_y(\frM)=\sum_{i=0}^d\, (-y)^i \ \sum_{j=0}^d\, (-1)^j\, \dim
H^j\big(\frM\,,\, \mbox{$\bigwedge^i$}\, T^*\frM\big)
\eea
is the Hirzebruch $\chi_y$-genus of the moduli space 
$\frM=\frM_{n,c}(\cO(-p))$ of $U(N)$ instantons on the surface
$\cO(-p)\to \IP^1$ of topological charge $n$ and magnetic charge $c$; here
$d=\dim_\IC\frM$. The D0 and D2 brane chemical potentials $\phi_0$ and
$\phi_2$ are related to the equivariant parameters of the
$\Omega$-deformation by
\bea
\phi_0= \frac{4\pi^2}{\epsilon_1} \qquad \mbox{and} \qquad \phi_2=
\frac{2\, \pi\, \theta}{\epsilon_1} \ ,
\eea
where here $\theta$ is interpreted as the four-dimensional
theta-angle, while
\bea
y=\e^{-2\pi\ii (\beta-1)} \ .
\eea
In the unrefined limit $\beta=1$, $y=1$, this is just the Vafa-Witten
partition function which is the generating function for the Euler
characteristics of instanton moduli spaces.

Following~\cite{Griguolo2006}, we use our formalism to derive conjectural
formulas for the $\chi_y$-genus of the surfaces $\cO(-p)\to \IP^1$,
which to the best of our knowledge are not known in closed form beyond
the case $p=1$ (where the geometry is simply that of $\IC^2$ blown up
at a point). For this, we keep only the classical contribution from
the refined Chern-Simons partition function (\ref{ZNCSbetasemiclass}), which
is associated to the boundary contribution to the four-dimensional instanton action,
and drop the perturbative contribution represented by the sum over the
Weyl group which should be absent from the partition function of the
topologically twisted $\cN=4$ gauge theory on $D$. This modifies
(\ref{ZqtQptorsion}) to the partition function
\beq
\widetilde{\cZ}_N^{\, {\rm BH}}(\phi_0,\phi_2,y;p) = \sum_{r\in\IZ_p^N} \
\sum_{n\in\IZ^N} \, \e^{-\phi_0\, \frac{(r+p\, n,r+p\,
    n)}{2p}-\phi_2\,\frac{|r+p\, n|}p}\,
y^{(n,\rho)} \ .
\label{cZBHbetap}\eeq
For $p=1$, this expression agrees with the contributions from fractional instantons to the anticipated generating function for the
$\chi_y$-genus of the instanton moduli space on $\cO(-1)\to\IP^1$, as
discussed in~\cite[\S5.5]{Aganagic2012}. For $p>1$, we conjecture that it is
the corresponding generating function for the surface
$\cO(-p)\to\IP^1$. We see explicitly from (\ref{cZBHbetap}) that the refinement keeps track of the contributions from each topological sector of fractional
instantons with fixed holonomy $r\in\IZ_p^N$ 
at infinity from
the finite action requirement that the gauge fields be asymptotically
flat; in the unrefined limit $y=1$, it can be resummed over $m=r+p\,
n\in\IZ^N$ to give the $N$-th power of the Jacobi theta-function
(\ref{Jacobitheta}) which
computes the fractional instanton contributions to the usual $\cN=4$ gauge
theory partition function on $\cO(-p)\to\IP^1$~\cite{Griguolo2006}.

\section{Planar limit\label{Planar}}

\subsection{Large $N$ limit of $(q,t)$-deformed Yang-Mills theory}

In this section we take the $N\to\infty$ limit of the refined $q$-deformed Yang-Mills partition
function (\ref{ZqtQpR}); we set the theta-angle equal to zero from now
on, i.e. $Q=1$. For this, we introduce the 't~Hooft
parameters $\tau_1$ and $\tau_2$ which are related to the deformation
parameters $\epsilon_{1}$ and $\epsilon_2$ as
\be
\tau_{1} = \epsilon_{1}\, N \qquad \mbox{and} \qquad
\tau_2=\epsilon_2\, N \ , 
\ee
and we keep these couplings large but fixed when taking $N$ large. In
this limit the refinement parameter $\beta={\epsilon_{2} \over \epsilon_{1}}=\frac{\tau_2}{\tau_1}$
is also kept fixed. We introduce the continuous distribution $R(x)$,
$x\in(0,1]$, of
partitions as
\be
R(x) = {R_i \over N} \qquad \mbox{for} \quad x = {i \over N} \ ,
\ee
which obeys $R(x)\geq R(y)$ for $x\leq y$, and the shifted distribution
\be
h(x) = -{R(x) \over \beta} + x - \frac{1}{2}
\ee
which obeys $h(x)\leq h(y)$ for $x\leq y$ and
\beq
h(x)-h(y)\geq x-y
\eeq
for $x\geq y$.

Writing the partition function (\ref{ZqtQpR}) as
\beq
Z(q,t,Q;p) = \sum_R\, \e^{-N^2\, S_R(\tau_1,\tau_2;p)} \ ,
\eeq
we find for the planar free energy
\bea
S_R(\tau_1,\tau_2;p) &=&- \beta \, \Big(\, \int_0^1\, \dd x \  \int_0^1\,  \dd
y \ \log \Big|2 \sinh \frac{\tau_1 \, \beta}{2} \, \big(h(x) - h(y)\big)\Big| +
\frac{p\, \tau_1 \, \beta}{2}\, \int_0^1\, \dd x \ h(x)^2  \nonumber \\
&& \qquad  -\, \frac{p \, \tau_1 \, \beta}{24} + \frac{2}{(\tau_1\,
  \beta)^2} \, F_0^{\rm{CS}} (\tau_1\, \beta) \, \Big) \ ,
\label{qtfreeenergy}\eea
where the line $x=y$ is excluded from the domain of the double integral;
we have used the fact that since $\beta$ is finite, the sum over $m$
coming from the dimension factors (\ref{qtdim}) is also finite and
thus $\frac mN\to0$ in the planar limit. 
Here
\be
\frac{2}{t^2} \, F_0^{\rm{CS}} (t) =
\int_0^1\, \dd x \ \int_0^1 \, \dd y \ \log \Big|2 \sinh \frac{t}{2}\, (x -y) \Big|
\ee
comes from the normalization factor (\ref{S00}), and it coincides with
the planar free energy of Chern-Simons theory on $S^3$ with 't~Hooft
coupling $t$; it can be also expanded as
\be
F_0^{\rm{CS}} (t) = \frac{t^3}{12} -
\frac{\pi^2 \, t}{6}
- {\rm Li}_3\big(\e^{-t}\big) + \zeta(3) \ , \label{CSfree}
\ee
where
\beq
{\rm Li}_3(x)=\sum_{n=1}^\infty \, \frac{x^n}{n^3}
\eeq
is the polylogarithm function of order $3$.
Apart from an overall factor of $\beta$, the $(q,t)$-deformed planar
Yang-Mills free energy (\ref{qtfreeenergy}) is related to the $q$-deformed free
energy $S_R(t;p)$ of~\cite{Arsiwalla:2005jb} by the simple change of the 't Hooft coupling
$t  =\tau_1\, \beta  = \tau_2$, where $t = g_s \, N$ is the
't Hooft coupling of the unrefined Yang-Mills theory,
i.e. $S_R(\tau_1,\tau_2;p)= \beta\, S_R(t=\tau_2;p)$; the unrefined limit
itself is of course obtained by setting $\beta =1$.

\subsection{Phase transition}

Using the simple relation between the unrefined and refined gauge
theories in the planar limit, we can easily write down the one-cut solution and the corresponding density of eigenvalues.
The saddle-point equation for the extrema $h(x)$ of the free energy
(\ref{qtfreeenergy}) is
\be
p \, h = \Pint\, \dd h' \ \rho(h'\, ) \, \coth\Big(\, \frac{\tau_1\,
  \beta}{2}\, (h-h'\,)\, \Big) \ ,
\label{saddlepointeq}\ee
where the spectral density
\beq
\rho(h):=\frac{\dd x}{\dd h}
\eeq
is bounded as
\beq
0< \rho(h)\leq 1
\label{rhohleq1}\eeq
and is normalized as $\int\, \dd h\ \rho(h)=1$. This principal value integral equation coincides with the large $N$ saddle-point equation of the Chern-Simons matrix
model \cite{Aganagic:2002wv,Marino:2004eq} with 't Hooft coupling $\tau_2$. 
More importantly, the planar limit of the $\beta$-deformed matrix
model for refined Chern-Simons theory on the three-sphere~\cite{Aganagic2011}
\bea
Z_N^{{\rm CS},\beta}(g_s) =\int_{\IR^N} \ \prod_{i=1}^N\, \dd u_i\
\e^{-\frac{u_i^2}{2g_s}} \
\prod_{m=0}^{\beta-1}\ \prod_{i\neq j} \, \big(\e^{(u_i-u_j)/2} -
q^m\, \e^{(u_j-u_i)/2}\big) 
\eea
gives the same saddle-point equation. 

Following precisely the same steps as in \cite{Arsiwalla:2005jb} we obtain for the density functional
\be
\rho(h) = \frac{p}{\pi} \, \arctan\bigg(\, \frac{\sqrt{
    \e^{A/p^2}-\cosh^2\big(\frac{A \, h}{2 p}\big)}}{\cosh\big(\frac{A
    \, h}{2 p}\big)}\, \bigg)
\ee
with the area parameter
\be
A := \tau_1 \, \beta \, p = \tau_2\, p \ .
\ee
The support of the spectral density is therefore $|h|<\frac{2p}A\,
{\rm arccosh}\big(\e^{A/2p^2}\big)$ and its range is
\be
{\rm im}(\rho) = \big[-\mbox{$\frac p2\,,\,\frac p2$} \, \big] \ .
\ee
Thus from (\ref{rhohleq1}) it follows that, similarly to the unrefined case, there
is no phase transition for $p \leq 2$. For $p > 2$, a phase transition
occurs when the density reaches its maximum value $1$, which is on the critical line
\be
A_*(p)=p^2\,\log\Big(\sec^2\big(\,\mbox{$\frac\pi p$}\,\big)\Big) \ .
\label{Apcritical}\ee
Following~\cite{Arsiwalla:2005jb} we can also write the refined
Yang-Mills free energy (\ref{qtfreeenergy}) in the small area phase as
\be
S_R(\tau_1, \tau_2; p) = \beta \, \Big(\, {1 \over \tau_2^2} \, \big(
p^2 \, F_0^{{\rm CS}} (\mbox{$\frac{\tau_2}p$} ) - 2 F_0^{{\rm CS}}
(\tau_2) \big) + {\tau_2 \over 12 p} + { p \, \tau_ 2 \over 24} 
\, \Big) \ , \label{freeenergy}
\ee
where $F_0^{{\rm CS}}(t)$ is the planar Chern-Simons free energy
(\ref{CSfree}).
Note that in the limit $p \rightarrow \infty$ one has
\bea
A_*(p) \ \longrightarrow \ \pi^2 \qquad \mbox{and} \qquad \rho(h) \
\longrightarrow \ \rho_G \big(h, A^{-1}\big) := \mbox{${A \over 2 \pi}$} \, \sqrt{{4 A^{-1}}  - h^2} \ ,
\eea
where $\rho_G (h, A)$ is the Wigner semicircle distribution of
the Gaussian matrix model which governs the planar small area phase of
ordinary Yang-Mills theory on $S^2$.

\subsection{Instanton contributions}

We now consider the large $N$ phase transition from the point of view
of the instanton expansion (\ref{ZqtQpm}) of the $(q,t)$-deformed
gauge theory. For this, we use the matrix integral representation
(\ref{cWm}) of the fluctuation factors to suitably perform the
$N\to\infty$ limit. In the planar limit, the measure factor
(\ref{hatDeltaqt}) simplifies drastically. Firstly, the factors
$\e^{\frac{\tau_1}{p\, N}\, (\mu,\mu+2\beta\, w(\rho))}\to1$ as
$N\to\infty$, since the refinement parameter $\beta$ is of order $1$ in the
limit and hence so are all root vectors $\mu\in\Lambda_\beta$; whence
$\widehat{\Delta}_{q,t}(x)\to \widetilde{\Delta}_{q,t}(x)$. Secondly,
in the Macdonald measure (\ref{Macmeasure}) one has
$q^m=\e^{-\frac{m\,\tau_1}N}\to1$ as $N\to\infty$; whence
$\Delta_{q,t}(x)\to\Delta(x)^\beta$ as before. Altogether we get
$\widehat{\Delta}_{q,t}(x)\to \Delta(x)^\beta\, \Delta(-x)^{\beta-1}$,
and defining $y_i=2\pi\,u_i$ for $i=1,\dots,N$ we can write the
fluctuation integral (\ref{cWm}) up to overall normalization in the planar limit as
\beq
\cW_{q,t}^\infty(p;m) = \int_{\IR^N}\ \prod_{i=1}^N\, \dd y_i\
\e^{-\frac{N\,\beta}{2A}\, y_i^2} \ \prod_{i<j}\,
\sin\Big(\, \frac{\tau_1\,\beta}{2A}\,\big(y_{ij}-2\pi\,
m_{ij}\big)\, \Big)\, \sin\Big(\,
\frac{\tau_1\,\beta}{2A}\,\big(y_{ij} +2\pi\,
m_{ij}) \Big)^{2\beta-1}
\label{cWmplanarm}\eeq
where we denote $x_{ij}:=x_i-x_j$.

Following~\cite{Arsiwalla:2005jb,Caporaso2005}, we look for a region
in parameter space where the one-instanton contribution dominates the zero-instanton
sector. Hence we define the function $\gamma(A,p)$ which measures the
relative weight of these contributions in the $N\to\infty$ limit
by
\beq
\exp\Big(-\frac{N\,\beta}{A}\, \gamma(A,p)\Big) = \e^{-\frac{2\pi^2\, N\,
  \beta} A} \
  \frac{\cW_{q,t}^\infty(p;e_i)}{\cW_{q,t}^\infty(p;0)}
  \ ,
\eeq
where $e_i\in\IZ^N$ is the vector with $1$ in its $i$-th entry and
$0$ in all other components; at $N=\infty$ the precise choice of
$e_i$ is immaterial.

The partition function describing the zero-instanton sector of the
gauge theory is defined by the $\beta$-deformed matrix integral
\beq
\cW_{q,t}^\infty(p;0) = \int_{\IR^N}\ \prod_{i=1}^N\, \dd y_i\
\e^{-\frac{N\,\beta}{2A}\, y_i^2} \ \prod_{i<j} \,
\sin\Big(\,
\frac{\tau_1\,\beta}{2A}\,\big(y_{i}-y_{j})\Big)^{2\beta} \ .
\label{cWmplanar}\eeq
The large $N$ limit is dominated by solutions of the saddle-point
equation
\beq
y=\tau_1\, \beta\, \Pint \, \dd y'\ \rho_{\rm inst}(y'\,)\, \cot\Big(\,
\frac{\tau_1\, \beta}{2A}\, \big(y-y'\,\big) \Big)
\eeq
for a suitable spectral distribution $\rho_{\rm inst}(y)$ for the
matrix model (\ref{cWmplanar}).
This equation is identical to the analogous saddle-point equation obtained
for the unrefined case in~\cite{Arsiwalla:2005jb,Caporaso2005}, and
hence we can simply read off the solution by substituting $t=\tau_1\,
\beta=\tau_2$ in their formulas. In particular, the spectral density is given
by
\beq
\rho_{\rm inst}(y) = \frac p{\pi\,A}\, {\rm arccosh}\Big(\e^{A/2p^2}\, \cos\Big(\,
\frac y{2p}\,\Big)\Big)
\label{rhoinsty}\eeq
with support $|y|<2p\,{\rm arccos}\big(\e^{-A/2p^2}\big)$. In the
limit $p\to\infty$, the density $\rho_{\rm inst}(y)\to
\rho_G(y,A)$ is the Wigner semicircle distribution with area
$A=\tau_1\, \beta\, p=\tau_2\,p$.

In the large $N$ limit, the function $\gamma(A,p)$ is completely
determined by the distribution (\ref{rhoinsty}) as
\bea
&& \exp\Big(-\, \frac{N\,\beta}{A}\, \gamma(A,p)\Big) \\ && = \int\, \dd y\
\exp\bigg(-\frac{N\, \beta}{2A}\, y^2+N\,\int\, \dd y'\ \rho_{\rm
  inst}(y'\,)\, \log\Big(\,\mbox{$
\frac{\sin\big(\,\frac{\tau_1\,\beta}{2A}\, (y-y'-2\pi)\,\big)\,
  \sin\big(\,\frac{\tau_1\,\beta}{2A}\,
  (y-y'+2\pi)\,\big)^{2\beta-1}}{\sin\big(\,\frac{\tau_1\,\beta}{2A}\,
  (y-y'\,)\,\big)^{2\beta}}$}\,\Big)\bigg) \nonumber 
\eea
and the integral over $y$ can be evaluated in the saddle-point
approximation by assuming that it is sharply peaked around $y=0$. Then we straightforwardly
obtain
\beq
\gamma(A,p)=2A\, \big(\cG(0)-\cG(2\pi)\big) \ ,
\label{gammaApfinal}\eeq
where we have defined the function
\beq
\cG(y):= \int\, \dd y'\ \rho_{\rm inst}(y'\,)\, \log\bigg|\sin\Big(\,
\frac{\tau_1\,\beta}{2A}\, (y-y'\,)\, \Big)\bigg|
\eeq
and used reflection symmetry $\cG(y)=\cG(-y)$. This function is
identical to that obtained in~\cite{Arsiwalla:2005jb,Caporaso2005} for
the unrefined case, and hence we can simply copy their solution with
the substitution $t=\tau_1\, \beta=\tau_2$ as before;
see~\cite[eq.~(4.22)]{Arsiwalla:2005jb} for the explicit form of the
function (\ref{gammaApfinal}). In
particular, we obtain in this way the standard critical area curve (\ref{Apcritical})
such that $\gamma(A_*(p),p)=0$; the instanton contributions to the
gauge theory partition function are exponentially suppressed for
$A<A_*(p)$ for all $p$, while at $A=A_*(p)$ the suppression ceases and
they become the favourable configurations. Hence just as in
the unrefined cases, the phase transition here is triggered by
two-dimensional instantons.

\section{Refined topological string theory} \label{geometry}

In this final section we discuss how the planar limit of the refined
gauge theories is related to refined topological string theory. For
this, we use the large $N$ duality between $U(N)$ Chern-Simons theory
on $S^3$ and topological string theory on the resolved conifold~\cite{Gopakumar:1998ki,Aganagic2011}.
The free energy computed in the weak coupling phase encodes
information about the emergent geometry. In fact, the
first piece of (\ref{freeenergy}), 
\beq
\beta \, {p^2 \over \tau_2^2} \, F_0^{{\rm CS}}
\big(\mbox{$\frac{\tau_2}p$} \big) \ , \label{CSpiece}
\eeq
which is the relevant term when comparing with the instanton
expansion, encodes the geometry of the resolved conifold with K\"ahler parameter $\frac{\tau_2}p$. As discussed in \cite{Aganagic2012},
for the refined topological string theory on the resolved conifold the
partition function can be explicitly computed. Its degree zero parts
are given by
\beq
\cZ_{0}^{\rm top} (q,t;\kappa) = \big( M(q, t) \, M(t, q) \big)^{\chi/4} \,
\e^{{1 \over \epsilon_1 \, \epsilon_2} \, {a \, \kappa^3 \over 6} + \beta\,
  {b \, \pi^2 \kappa \over 6}}
\label{Z0topqt}\eeq
where the K\"ahler parameter is $\kappa=\frac{\tau_2}p$ in our case, and 
\beq
M(q,t) = \prod_{n,m=1}^{\infty}\, \big( 1 - t^n\, q^{m-1} \big)  
\eeq
is the refined MacMahon function. Here $\chi = 2$ is the Euler characteristic of the conifold, while $a$ and $b$ are constants which are
related to the triple intersection product of the K\"ahler class and to the
second Chern class of the Calabi-Yau manifold, respectively. Since our manifold is non-compact,
the choices for these constants are ambiguous; we choose them so
that in the unrefined limit the expression (\ref{Z0topqt}) agrees with the usual
free energy of the conifold \cite{Gopakumar:1998ki}. Then the contribution to the genus zero free energy of the conifold is
\beq
\cF_0^{\rm top}(q,t;\kappa) = \beta \, \Big( \zeta(3)  + {\kappa^3 \over
  12} - {\pi^2 \, \kappa \over 6} \Big) \ .
\eeq 
The non-trivial and unambiguous contribution to the partition function can be computed for example from the refined topological vertex
\cite{Iqbal:2007ii}, and is given by
\beq
\cZ^{{\rm top}}(q,t;\kappa) = \exp\Big(- \sum_{n=1}^{\infty} \,
{\kappa^n \over n \, \big(q^{n/2} - q^{-n/2}\big)\, \big(t^{n/2} -
  t^{-n/2} \big)} \Big) \ .
\eeq
Fixing $g_s = \epsilon_2$, this gives the additional genus
zero contribution $-\beta\, {\rm Li}_3(\e^{-\kappa})$, so that the
total genus zero free 
energy of closed refined topological string theory on the conifold is
given by
\beq
F_0^{{\rm con}}(q,t;\kappa) = \beta\, \Big( \zeta(3)  + {1 \over 12}
\, \Big(\, {\tau_2 \over p }\, \Big)^3 - {\pi^2 \over 6}\, {\tau_2
  \over p } - {\rm Li}_3 \big(\e^{- {\tau_2 / p }} \big) \Big) \ .
\eeq
This expression agrees with the Chern-Simons contribution (\ref{CSpiece}); the
agreement is due to the geometric transition, and is related to the emergent conifold geometry seen in
the strong coupling instantonic phase.

The reason why the conifold geometry emerges in the weak coupling
phase can be understood by noting that in this
phase the geometry is described by the $\beta$-deformed matrix model (\ref{cWmplanar})
of the zero-instanton sector; in the refined Chern-Simons description
this corresponds to the contribution from the trivial flat
connection. In fact, the full weak coupling geometry can be equivalently found from the planar limit of
the corresponding Chern-Simons matrix model on $S^3$ as in
\cite[eq.~(4.5)]{Aganagic:2002wv} with the K\"ahler parameter
$t=\frac{\tau_2}p$,
where it is also shown that its mirror geometry precisely describes the
conifold. It is exactly this mirror conifold that one sees in the
weak coupling phase of the gauge theory. 

The free energy (\ref{freeenergy}) also contains additional terms
which do not appear to admit such an interpretation in terms of
refined topological string theory. However, we must remember that,
like in the unrefined case, the weak coupling phase is not expected to
yield the correct description of the large $N$ dual geometry. 
In the strong coupling phase multi-instantons contribute
and the geometry is controlled by the more general matrix integrals
(\ref{cWmplanar}) evaluated on torsion vectors $r\in\IZ_p^N$
parametrizing non-trivial flat connections. We expect that the
inclusion of all flat connections will restore the anticipated dual
geometry to the cotangent bundle $T^*(S^3/\IZ_p)$, which is an $A_{p-1}$ fibration over
$\IP^1$~\cite{Aganagic:2002wv}. For this, one should construct the
two-cut solution of the saddle-point equation (\ref{saddlepointeq})
appropriate to the large area phase; we have refrained from attempting
this, as even in the unrefined case only partial results are available
from the two-cut analysis~\cite{Arsiwalla:2005jb,Caporaso2005}. Note
that the fluctuation integrals $\cW_{q,t}(p;r)$ from (\ref{cWm}) for
$r\in\IZ_p^N$ have a natural interpretation in terms of a fixed
configuration of $N$ topological
D3-branes wrapped on the cycles of $S^3/\IZ_p$; however, the refined matrix
model is sensitive to their insertion points because of the Weyl
symmetry breaking, which disappears in the planar limit. It would be
interesting to understand further the closed topological string theory
emerging from the geometric transition through the matrix model
geometry determined by (\ref{cWm}).

As the refined conifold partition function emerges for any value of
the degree $p$, it is natural to ask what becomes of the $p\to\infty$
limit. As we have seen, in this limit the small area phase is governed
by the $\beta$-deformed Gaussian matrix model
\beq
\cZ_\infty^{{\rm YM},\beta}(a) = \int_{\IR^N} \ \prod_{i=1}^N\, \dd
u_i \ \e^{-\frac a2\, u_i^2} \ \prod_{i<j}\, (u_i-u_j)^{2\beta} \ .
\eeq
In the large $N$ limit, it is shown in~\cite{DijkgraafVafa2009} that
this matrix integral corresponds to the refined conifold geometry and
coincides with the partition function of two-dimensional $c=1$ string
theory at radius $R=\beta$; $\beta$-deformed
matrix ensembles are also used in~\cite{Aganagic:2011mi} to discuss
refined topological string theory. This result is in full agreement with the analysis of the
planar limit of refined Chern-Simons theory on $S^3$ carried out
in~\cite{Krefl:2013gua}, where it is shown explicitly that the
refinement replaces the virtual Euler characteristic of the moduli
space of complex curves (which computes the perturbative free energy
of the ordinary topological string theory on the resolved
conifold~\cite{Gopakumar:1998ki}) with a parametrized Euler
characteristic appropriate to the radius deformed $c=1$ string theory.

\section*{Acknowledgments}

R.J.S. thanks the staff of the Institute of Theoretical Physics at
E\"otv\"os Lor\'and University for the warm hospitality during the final
stages of this work. The work of R.J.S. was supported in part by the
Consolidated Grant ST/J000310/1 from the UK Science and Technology
Facilities Council, and by Grant RPG-404 from the Leverhulme Trust.

\appendix

\section{Generalized Weyl denominator formula}

A generalization of the Weyl character formula to Macdonald
polynomials, seen as quantum group $\beta$-deformations of Schur polynomials, has
been developed
in~\cite{Felder,EtingofStyrkas,Chalykh,ChalykhEtingof}. By considering
the Macdonald polynomial corresponding to the trivial representation,
which is equal to $1$, we can extract a generalization of the Weyl
denominator identity. In particular,
from~\cite[Theorem~3.11]{ChalykhEtingof} we infer the identity
\beq
\widetilde{\Delta}_{q,t}(x):=\frac{\Delta_{q,t}(x)\,
  \Delta_{q,t}(-x)}{\Delta(-x)} = \sum_{w\in S_N}\, \varepsilon(w)\,
\Psi_{\beta\,w(\rho)}(x;q,t)
\label{genWeylid}\eeq
where $\varepsilon(w)$ is the sign of the Weyl group element $w\in S_N$ which
acts on $\IC^N$ by permutating components of $N$-vectors. The meromorphic function
$\Psi_\lambda(x;q,t)$ of $q$ and the $U(N)$ weights $\lambda$ is a complicated
Laurent series. An explicit but
involved combinatorial expansion can be found in~\cite[\S8]{Felder};
in~\cite[\S5]{EtingofStyrkas} it is described in terms of generalised
characters, while in~\cite{Chalykh,ChalykhEtingof} it is called a
normalised Baker-Akhiezer function and constructed via applications of
Macdonald difference operators to the function $\widetilde{\Delta}_{q,t}(x)$. Its main characteristics can be
summarised as follows. Let us define the set
\beq
\Lambda_\beta^\circ :=\Big\{\mu=\sum_{\alpha>0}\, \mu_\alpha\,\alpha \ \Big| \
0\leq \mu_\alpha\leq \beta-1\Big\} \ ,
\eeq
where the sums run over positive roots of the Lie algebra of the unitary group $U(N)$. We
may parametrize elements $\mu\in\Lambda^\circ_\beta$ by sequences of integers
$\mu=\{\mu_{ij}\}_{1\leq i<j\leq N}$ with $0\leq\mu_{ij}\leq \beta-1$
and $(\mu,x)=\sum_{i<j}\, \mu_{ij}\, (x_i-x_j)$. Then for
$t=q^{1-\beta}$, the function
$\Psi_\lambda(x;q,t)$ can be expanded in the form
\beq
\Psi_\lambda(x;q,t) = \e^{(\lambda,x)} \ \sum_{\mu\in\Lambda^\circ_\beta}\, \e^{(\mu,x)}\
\Pi_\mu(\lambda;q,t) \ ,
\label{Psiexp}\eeq
where the expansion coefficients $\Pi_\mu(\lambda;q,t)$ are normalised such that
\beq
\Pi_0(\lambda;q,t)=1 \ .
\label{Pi0norm}\eeq
It has the following properties:
\begin{itemize}
\item $\Psi_{w(\lambda)}\big(w(x);q,t\big)=\Psi_\lambda(x;q,t)$ for all $w\in S_N$.
\item $\Psi_{-\lambda}(-x;q,t)=\Psi_\lambda(x;q,t)$.
\item $\Psi_\lambda(x;q^{-1},t)=\Psi_\lambda(-x;q,t)$.
\end{itemize}
In the case of interest in this paper, wherein $t=q^\beta$, one must
understand the expansion (\ref{Psiexp}) as an analytical continuation
by replacing $\Lambda_\beta^\circ$ with an infinite subset $\Lambda_\beta$
of the root lattice of $U(N)$. In this case the series in
(\ref{Psiexp}) is infinite but still given by elementary functions, and
it possesses the same properties as those listed above;
see~\cite[\S3.5]{ChalykhEtingof}, \cite[\S3]{EtingofStyrkas}
and~\cite[\S8]{Felder} for details.

In the unrefined limit $\beta=1$, only the $\mu=0$ contribution remains
of the sum in (\ref{Psiexp}). By (\ref{Pi0norm}), in this case
\beq
\Psi_\lambda(x;q,q) = \e^{(\lambda,x)}
\eeq
is the usual character of the Verma module $\cM_\lambda$ for $U(N)$, and the expansion (\ref{genWeylid}) reduces to the usual Weyl denominator formula
\beq
\Delta(x) = \widetilde{\Delta}_{q,q}(x) = \sum_{w\in S_N}\,
\varepsilon(w)\, \e^{(w(\rho),x)} \ .
\eeq

\end{document}